\input psfig

\documentstyle{article}

\begin{document}

\vspace{1 cm}
\Large
\title{\bf Partitioning of a polymer chain between two confining cavities: 
the role of electrostatic interactions}

\vspace{1.8 cm}
\large
\author{Stefan Tsonchev and Rob D. Coalson \\ 
Department of Chemistry, University of Pittsburgh, Pittsburgh, PA 15260
\and
Anthony Duncan \\ Department of Physics, University of Pittsburgh, Pittsburgh, PA 15260}
\date{}
\maketitle

\begin{abstract}
A recently developed lattice field theory approach to the statistical 
mechanics of charged polymers in electrolyte solutions [S.~Tsonchev, 
R.~D.~Coalson, and A.~Duncan, Phys. Rev. E {\bf{60}}, 4257, (1999)] is 
generalized to the case where ground-state dominance in the polymer's Green's 
function does not apply. The full mean-field equations for the system are 
derived and are shown to possess a unique solution. The 
approach is applied to the problem of a charged Gaussian polymer chain 
confined to move within the region defined by two fused spheres. The failure 
of the notion of ground-state dominance under certain conditions even in 
the limit of large number of monomers is demonstrated.
\end{abstract}

\newpage
\section{Introduction}
The problem of partitioning of a polymer chain between cavities of different 
size has been of interest to researchers for some time \cite{Cas, CasTag, 
Mutu}. Investigations of this phenomenon are motivated by the practical 
importance of techniques for separation of macromolecules according to their 
size, such as gel electrophoresis, size exclusion chromatography, membrane 
separation, filtration, etc. \cite{Rod}. All these methods rely on the 
different mobility of macromolecules of different size as they migrate 
through a porous network of random obstacles. Such networks can be modeled as 
a complex system of interconnected cavities and channels available to the 
polymer chain. Hence, understanding how
a polymer partitions itself between such cavities can lead to more efficient 
separation methods. Ultimately, one hopes to utilize the 
dependence of polymer partitioning on molecular properties, such as polymer 
length, electrical charge and electrolyte composition, to 
selectively separate chains according to their molecular weight.

Recently, experimental and theoretical investigations \cite{Asher, Chern} 
have explored the so called ``entropic trapping'' 
of polymer chains in large spherical voids in a gel, and have opened the door 
to potential new methods of macromolecular separation. ``Entropic trapping'' 
describes the process of preferential localization of a polymer chain inside 
a void of larger size than the typical channels and voids in a gel, due to 
the larger conformational entropy experienced by the polymer in the large 
void as compared to the one in the narrow channels and smaller voids of the 
gel. The trapping of the polymer in the cavity is characterized by the 
partition coefficient $K$, which in the case of a gel is defined as the ratio 
of the polymer concentrations inside and outside the cavity. 

In a recent letter \cite{Ts} we investigated the role of excluded 
volume interactions between monomers on the partitioning of a polymer chain 
between
two connected spheres.  In particular, we determined the dependence of the 
partition coefficient $K$ (in this case, the ratio of the average
number of monomers in each of the
two spheres)
on the total number of monomers in 
the chain.
The results, which 
are in accord with recent 
experiments \cite{Asher} and computer simulations \cite{Chern}, showed
that current theoretical understanding of polymer partitioning is 
incomplete. It was also shown that for certain kinds of systems the notion of 
ground-state dominance of the polymer's Green's function can fail even in the 
limit of large number of monomers.

In this paper we investigate the role of electrostatic interactions on the 
partitioning of a charged polymer chain between two confining spherical
cavities
of different size. We have carried out calculations using the Lattice Field 
Theory (LFT) approach \cite{TCD}, based on a formalism which has a number of 
antecedents in the literature [9--14].

In Section 2 we generalize the LFT approach to the situation where 
ground-state dominance of the polymer's Green's function is not assumed.
Then in 
Section 3 we discuss the general shape of the total free energy functional 
of the system at the mean-field level, and prove that it has a unique minimum,
thus guaranteeing the 
existence of a unique solution of the mean-field equations. 
Numerical results from calculations using three-dimensional (3D) LFT 
are presented in Section 4, and in Section 5 we summarize our conclusions.

\newpage
\section{Lattice Field Theory of Charged Polymer Chains in Electrolyte 
Solution}
We first extend the formalism presented in \cite{TCD} to the general case 
where ground-state dominance of the polymer's Green's function does not apply.
In Ref. \cite{TCD} we showed that
the full partition function of a charged polymer in an 
electrolyte solution with short-range monomer repulsion interactions can be
written as a functional integral: 
\begin{equation}
 Z=\int 
D\chi(\vec{r})D\omega(\vec{r})e^{\frac{\beta\epsilon}{8\pi}\int\chi\Delta\chi 
d\vec{r}-\frac{\lambda}{2}\int\omega(\vec{r})^{2}d\vec{r}+c_{+}\int 
e^{ie\beta\chi}d\vec{r}+c_{-}\int e^{-ie\beta\chi}d\vec{r}}Z_{Schr}(\chi,\omega) \, ,
\label{Z}
\end{equation}
where $\beta{=}1/kT$ is the inverse temperature, $\epsilon$ is the dielectric 
constant of the solution, $e$ is the electron charge, $\lambda$ is a 
measure of the strength of the excluded volume interaction, 
$\chi$ and $\omega$ are auxiliary fields, 
$c_{\pm}{=}e^{\beta\mu_{\pm}}/\lambda_{\pm}^{3}$ with $\mu_{\pm}$ and 
$\lambda_{\pm}$ being the chemical potentials and the thermal deBroglie 
wavelengths for the ions, respectively, and 
\begin{equation}
Z_{Schr}(\chi,\omega)\equiv \int 
D\vec{x}(s)e^{-\frac{3}{2a_{p}^{2}}\int_{0}^{M}ds 
\dot{\vec{x}}^{2}(s)-ipe\beta\int  ds\chi(\vec{x}(s))-i\lambda\int 
ds\omega(\vec{x}(s))} \, ,
\label{ZSchr}
\end{equation}
with $M$ being the total number of monomers in the chain, $p$ the charge per 
monomer, and $a_{p}$ the Kuhn length.

As before \cite{TCD}, the functional integral (\ref{Z}) can be rerouted 
through a 
complex saddle-point at $\chi{=}{-}i\chi_{c}$ and $\omega{=}{-}i\omega_{c}$, 
where $\chi_{c}$ and $\omega_{c}$ are purely real, reducing the computation 
of $Z_{{\rm Schr}}(\chi,\omega)$ at the saddle-point to a conventional 3D 
Schr\"odinger Hamiltonian problem, that is, the computation of matrix 
elements of $e^{-HT}$, with Euclidean time extent of the evolution $T{=}M$ 
and 
\begin{equation}
  H\equiv -\frac{a_{p}^{2}}{6}\vec{\nabla}^{2}+\lambda\omega_{c}(\vec{r})
+\beta pe\chi_{c}(\vec{r}) \, .
\label{eq:H}
\end{equation}
As usual, the equations determining the saddle-point configuration fields 
$\chi_{c}$, $\omega_{c}$ are obtained by setting the variational derivative 
of the exponent in the full functional integral (\ref{Z}) to zero. In the 
general case of a polymer chain with free ends the polymer part of the 
partition function can be written:
\begin{eqnarray}
Z_{Schr}&=&{\int}dx_{i}dx_{f}\sum_{n}\Psi_{n}(x_{i})\Psi_{n}(x_{f})e^{-ME_{n}} \nonumber \\
&=&\sum_{n}A_{n}^{2}e^{-ME_{n}}\,{\equiv}\,e^{F_{pol}} \, ,
\label{eFpol}
\end{eqnarray}
where $E_{n}$ is the $n$-th energy eigenvalue,
\begin{equation}
A_{n}\equiv{\int}d\vec{r}\,\Psi_{n}(\vec{r}) \, ,
\label{An}
\end{equation}
and
\begin{equation}
F_{pol}=\ln\left(\sum_{n}A_{n}^{2}e^{-ME_{n}}\right) \, .
\label{Fpol}
\end{equation}
is the negative of the polymer contribution to the free energy.
Thus the negative of the total free energy at the saddle-point becomes
\begin{equation}
  F=\int d\vec{r}\left\lbrace 
\frac{\beta\epsilon}{8\pi}\left|\vec{\nabla}\chi_{c}\right|^{2}+\frac{\lambda}{2}
\omega_{c}^{2}+c_{+} e^{\beta e\chi_{c}}+c_{-} e^{-\beta e\chi_{c}}\right\rbrace
+F_{pol}(\chi_{c},\omega_{c}) \, .
\label{F}
\end{equation}
As before \cite{TCD}, we have
\begin{eqnarray}
    \frac{\delta E_{n}}{\delta\chi_{c}(\vec{r})} &=& \beta 
pe|\Psi_{n}(\vec{r})|^{2} \, , \\
    \frac{\delta E_{n}}{\delta\omega_{c}(\vec{r})} &=& 
\lambda|\Psi_{n}(\vec{r})|^{2} \, .
\label{varE}
\end{eqnarray}
Considering the variation of the potential energy $V$, 
${\delta}V{=}{\beta}pe\delta\chi_{c}$, we have
\begin{eqnarray}
\delta\Psi_{n}(\vec{r})&=&\sum_{m,m{\ne}n}\frac{\left<m|{\delta}V|n\right>}{E_{n}-E_{m}}\Psi_{m}(\vec{r}) \nonumber \\
&=&{\beta}pe\sum_{m,m{\ne}n}\frac{\Psi_{m}(\vec{r})}{E_{n}-E_{m}}{\int}d\vec{r}\,''\Psi_{m}(\vec{r}\,'')\Psi_{n}(\vec{r}\,''){\delta}\chi_{c}(\vec{r}\,'') \, ,
\label{varPsi}
\end{eqnarray}
so that
\begin{equation}
\frac{\delta\Psi_{n}(\vec{r}\,')}{\delta\chi_{c}(\vec{r})}={\beta}pe\sum_{m,m{\ne}n}\frac{\Psi_{m}(\vec{r}\,')\Psi_{m}(\vec{r})\Psi_{n}(\vec{r})}{E_{n}-E_{m}} \, .
\label{varPsi/varchi}
\end{equation}
Hence,
\begin{equation}
\frac{\delta}{\delta\chi_{c}(\vec{r})}\left(A_{n}^{2}e^{-ME_{n}}\right)=2{\beta}peA_{n}\Psi_{n}(\vec{r})\sum_{m,m{\ne}n}\frac{A_{m}\Psi_{m}(\vec{r})}{E_{n}-E_{m}}e^{-ME_{n}}-{\beta}peMA_{n}^{2}\Psi_{n}^{2}(\vec{r})e^{-ME_{n}} \, ,
\label{var/varchi}
\end{equation}
and
\begin{eqnarray}
\frac{{\delta}F_{pol}}{\delta\chi_{c}(\vec{r})}&=&\frac{\sum_{n}\frac{\delta}{\delta\chi_{c}(\vec{r})}\left(A_{n}^{2}e^{-ME_{n}}\right)}{\sum_{n}A_{n}^{2}e^{-ME_{n}}} \nonumber \\
&=&{\beta}pe\frac{\sum_{n,m,m{\ne}n}\frac{2A_{n}\Psi_{n}(\vec{r})A_{m}\Psi_{m}(\vec{r})}{E_{n}-E_{m}}e^{-ME_{n}}-M\sum_{n}\left(A_{n}\Psi_{n}(\vec{r})\right)^{2}e^{-ME_{n}}}{\sum_{n}A_{n}^{2}e^{-ME_{n}}} \, .
\label{varFpol/varchi}
\end{eqnarray}
Considering the fact that
\begin{eqnarray}
\sum_{n,m,m{\ne}n}\frac{A_{n}\Psi_{n}A_{m}\Psi_{m}}{E_{n}-E_{m}}e^{-ME_{n}}&=&\sum_{m,n,n{\ne}m}\frac{A_{n}\Psi_{n}A_{m}\Psi_{m}}{E_{m}-E_{n}}e^{-ME_{m}} \nonumber \\
&=&\frac{1}{2}\sum_{n,m,m{\ne}n}\frac{A_{n}\Psi_{n}A_{m}\Psi_{m}}{E_{n}-E_{m}}\left(e^{-ME_{n}}-e^{-ME_{m}}\right) \, ,
\label{sumMN}
\end{eqnarray}
and
\begin{equation}
\frac{A_{n}\Psi_{n}A_{m}\Psi_{m}}{E_{n}-E_{m}}\left(e^{-ME_{n}}-e^{-ME_{m}}\right)\stackrel{E_{n}{\rightarrow}E_{m}}{\longrightarrow}-M\left(A_{n}\Psi_{n}\right)^{2}e^{-ME_{n}} \, ,
\label{En=Em}
\end{equation}
we can write (\ref{varFpol/varchi}) as an unconstrained double sum over 
states:
\begin{equation}
\frac{{\delta}F_{pol}}{\delta\chi_{c}(\vec{r})}={\beta}pe\frac{\sum_{n,m}\frac{A_{n}\Psi_{n}A_{m}\Psi_{m}}{E_{n}-E_{m}}\left(e^{-ME_{n}}-e^{-ME_{m}}\right)}{\sum_{n}A_{n}^{2}e^{-ME_{n}}} \, .
\label{varF/varchi2}
\end{equation}
In a similar fashion we obtain:
\begin{equation}
\frac{{\delta}F_{pol}}{\delta\omega_{c}(\vec{r})}=\lambda\frac{\sum_{n,m}\frac{A_{n}\Psi_{n}A_{m}\Psi_{m}}{E_{n}-E_{m}}\left(e^{-ME_{n}}-e^{-ME_{m}}\right)}{\sum_{n}A_{n}^{2}e^{-ME_{n}}} \, .
\label{varF/varomega2}
\end{equation}
As shown below, the quantity
\begin{equation}
\rho(\vec{r}){\equiv}-\frac{\sum_{n,m}\frac{A_{n}\Psi_{n}A_{m}\Psi_{m}}{E_{n}-E_{m}}\left(e^{-ME_{n}}-e^{-ME_{m}}\right)}{\sum_{n}A_{n}^{2}e^{-ME_{n}}} 
\label{rho}
\end{equation}
is the total monomer density. Thus, we can finally write down the equations 
determining the saddle point configuration fields $\chi_{c}$, $\omega_{c}$:
\begin{eqnarray}
    \frac{1}{\beta e}\frac{\delta F}{\delta\chi_{c}(\vec{r})} &=& 
-\frac{\epsilon}{4\pi e}\vec{\nabla}^{2}\chi_{c}(\vec{r})+ c_{+}e^{\beta 
e\chi_{c}(\vec{r})}
 - c_{-}e^{-\beta e\chi_{c}(\vec{r})}-p\rho(\vec{r})=0 
\label{fe1} \, , \\
   \frac{1}{\lambda} \frac{\delta F}{\delta\omega_{c}(\vec{r})} 
&=&\omega_{c}(\vec{r})-\rho(\vec{r})=0 \, .
\label{fe2}
\end{eqnarray}
Using Eq. (\ref{fe2}) the auxiliary field $\omega_{c}(\vec{r})$ can 
be eliminated, leaving the following pair of coupled nonlinear equations
to describe, at mean-field level,
the equilibrium properties of a charged polymer 
interacting with ions in an electrolyte solution:
\begin{eqnarray}
  \frac{\epsilon}{4\pi e}\vec{\nabla}^{2}\chi_{c}(\vec{r})&=&c_{+}e^{\beta 
e\chi_{c}(\vec{r})}
 - c_{-}e^{-\beta e\chi_{c}(\vec{r})}-p\rho(\vec{r})
\label{eq:pb} \, ,  \\
  \frac{a_{p}^{2}}{6}\vec{\nabla}^{2}\Psi_{n}(\vec{r})&=&\lambda 
\rho(\vec{r})\Psi_{n}(\vec{r})
+\beta pe\chi_{c}(\vec{r})\Psi_{n}(\vec{r})-E_{n}\Psi_{n}(\vec{r}) \, .
\label{eq:nlse}
\end{eqnarray}
Eqs. (\ref{eq:pb}) and (\ref{eq:nlse}) are generalizations of 
equations (20) and (21) in 
\cite{TCD}, as they involve the total monomer density, $\rho(\vec{r})$, given 
by (\ref{rho}), instead of $M\Psi_{0}^{2}(\vec{r})$, 
which would be appropriate only in the limit of ground-state dominance.
The equations presented here apply
for polymer chains of arbitrary length. Inclusion 
of single-particle potentials, which can be used to enforce exclusion regions 
for either the ions or monomers is straightforward \cite{TCD}. It is 
important to note that the parameters $c_{\pm}$ are exponentials of the 
chemical potentials $\mu_{\pm}$ for positively and negatively charged ions, 
the numbers of these ions must be fixed by suitably adjusting $c_{\pm}$  
to satisfy the relations
\begin{equation}
  n_{\pm}=c_{\pm}\frac{\partial \log{(Z)}}{\partial c_{\pm}}=c_{\pm}
\int e^{\pm\beta e\chi_{c}}d\vec{r} \, .
\end{equation}

Next we show that $\rho(\vec{r})$ is in fact the 
monomer density. Starting from the average of the local monomer density, 
$\rho_{n}(\vec{r})$ \cite{Edwards}, we have
\begin{eqnarray}
\rho_{n}(\vec{r})&=&\left<\delta\left(\vec{r}-\vec{R}_{n}\right)\right>\nonumber \\
&=&\frac{{\int}d\vec{R}_{0}d\vec{R}_{M}\left<\vec{R}_{M}\left|e^{-(M-n)\hat{H}}\right|\vec{r}\right>\left<\vec{r}\left|e^{-n\hat{H}}\right|\vec{R}_{0}\right>}{{\int}d\vec{R}_{0}d\vec{R}_{M}\left<\vec{R}_{M}\left|e^{-M\hat{H}}\right|\vec{R}_{0}\right>} \, .
\end{eqnarray}
Then the total density is
\begin{eqnarray}
\rho(\vec{r})&=&\sum_{n=0}^{M}\rho_{n}(\vec{r})\,\approx\,\int_{0}^{M}dn\rho_{n}(\vec{r})\nonumber \\
&=&\frac{\int_{0}^{M}dn{\int}d\vec{R}_{0}d\vec{R}_{M}\left<\vec{R}_{M}\left|e^{-(M-n)\hat{H}}\right|\vec{r}\right>\left<\vec{r}\left|e^{-n\hat{H}}\right|\vec{R}_{0}\right>}{{\int}d\vec{R}_{0}d\vec{R}_{M}\left<\vec{R}_{M}\left|e^{-M\hat{H}}\right|\vec{R}_{0}\right>} \, .
\end{eqnarray}
Invoking spectral decomposition
\begin{eqnarray}
\left<\vec{r}_{2}\left|e^{-m\hat{H}}\right|\vec{r}_{1}\right>&=&\sum_{j=0}^{\infty}\left<\vec{r}_{2}|\Psi_{j}\right>\left<\Psi_{j}|\vec{r}_{1}\right>e^{-mE_{j}} \nonumber \\
&=&\sum_{j=0}^{\infty}\Psi_{j}(\vec{r}_{2})\Psi_{j}(\vec{r}_{1})e^{-mE_{j}} \, ,
\label{spd}
\end{eqnarray}
we obtain 
\begin{eqnarray}
{\int}d\vec{R}_{0}d\vec{R}_{M}\left<\vec{R}_{M}\left|e^{-M\hat{H}}\right|\vec{R}_{0}\right>&=&\sum_{j=0}^{\infty}{\int}d\vec{R}_{0}d\vec{R}_{M}\Psi_{j}(\vec{R}_{0})\Psi_{j}(\vec{R}_{M})e^{-ME_{j}} \nonumber \\
&=&\sum_{j=0}^{\infty}A_{j}^{2}e^{-ME_{j}} \, .
\end{eqnarray}
Letting
\begin{eqnarray}
I(n)&=&{\int}d\vec{R}_{0}d\vec{R}_{M}\left<\vec{R}_{M}\left|e^{-(M-n)\hat{H}}\right|\vec{r}\right>\left<\vec{r}\left|e^{-n\hat{H}}\right|\vec{R}_{0}\right> \nonumber \\
&=&{\int}d\vec{R}_{0}d\vec{R}_{M}\left[\sum_{j=0}^{\infty}\Psi_{j}(\vec{R}_{M})\Psi_{j}(\vec{r})e^{-(M-n)E_{j}}\right]\left[\sum_{k=0}^{\infty}\Psi_{k}(\vec{R}_{0})\Psi_{k}(\vec{r})e^{-nE_{k}}\right] \nonumber \\
&=&\sum_{j=0}^{\infty}\sum_{k=0}^{\infty}A_{j}A_{k}\Psi_{j}\Psi_{k}e^{-ME_{j}}e^{-n(E_{k}-E_{j})} \, ,
\end{eqnarray}
then
\begin{eqnarray}
\int_{0}^{M}dnI(n)&=&\sum_{j=0}^{\infty}\sum_{k=0}^{\infty}A_{j}A_{k}\Psi_{j}\Psi_{k}e^{-ME_{j}}\int_{0}^{M}dne^{-n(E_{k}-E_{j})} \nonumber \\
&=&\sum_{j=0}^{\infty}\sum_{k=0}^{\infty}A_{j}A_{k}\Psi_{j}\Psi_{k}f(M;E_{j},E_{k}) \, ,
\end{eqnarray}
where
\[ f(M;E_{j},E_{k})=\left\{ \begin{array}{cc} 
\frac{e^{-ME_{j}}-e^{-ME_{k}}}{E_{k}-E_{j}}\,\,\,\,\,; \,for \,\,E_{j} \neq E_{k}&\\
Me^{-ME_{j}} \,\,\,\,\,\,\,\,\,\,\,\,\,\,\,; \,for \,\,E_{j} = E_{k}&
\end{array} \right. \, . \]
Thus, all in all
\begin{equation}
\rho(\vec{r})=\frac{\sum_{j=0}^{\infty}\sum_{k=0}^{\infty}A_{j}A_{k}\Psi_{j}\Psi_{k}f(M;E_{j},E_{k})}{\sum_{j=0}^{\infty}A_{j}^{2}e^{-ME_{j}}} \, ,
\end{equation}
thereby recovering Eq. (\ref{rho}).
It is easy to see that in the case of ground-state dominance, that is, if in 
the spectral decomposition formula (\ref{spd}) 
\begin{equation}
M(E_{1}-E_{0})>>1 \, ,
\end{equation}
we recover the familiar relationship
\begin{equation}
\rho(\vec{r})=M\Psi_{0}^{2}(\vec{r}) \, ,
\end{equation}
and equations (\ref{eq:pb}) and (\ref{eq:nlse}) become identical with 
equations (20) and (21) in \cite{TCD}.

In concluding this section, we should mention that the functional
$F(\chi_c, \omega_c)$ in Eq. (\ref{F}) represents a ``mixed"
free energy, since the partition function $Z$ in Eq. (\ref{Z})
is canonical with respect to the polymer chain and grand
canonical with respect to the mobile ions.  The advantage
of working with $F$ is that, as will be shown in the next section, it has a 
unique minimum (corresponding to the solution of the mean
field Eqs. (\ref{eq:pb}) and (\ref{eq:nlse})), and thus can be used to guide
a numerical search for the mean electrostatic and monomer
density fields.  Once the mean fields have been computed,
the defining relation ${\ln}Z{\cong}F(\chi_c,\omega_c)$ can
be used to obtain free energies of various types.
For example, the Helmholtz free energy $A$ (corresponding
to fixed numbers of monomers and impurity ions) is given by
\begin{equation}
\beta A = n_+ \ln c_+ + n_- \ln c_- - F(\chi_c,\omega_c) \, .
\end{equation}

\newpage
\section{General Form of the Free Energy Functional}
In our previous work \cite{TCD} we showed that in the case of ground-state 
dominance the negative of the free energy, $F$, is in fact convex, thus 
guaranteeing a unique 
solution to the field equations. Of course, convexity, although a sufficient 
condition, is not necessary for the existence of a unique local minimum of 
$F$. We shall see shortly that a weaker, but nevertheless
 sufficient condition for the existence of a unique minimum of $F$  is the 
convexity of the functional $e^{F}$.  This latter convexity will be 
established below for arbitrary boundary conditions  (open or closed) for the 
polymer chain, implying a unique minimum of $F$---a property essential for the 
stability and reliability of the numerical algorithms we employ to solve the 
mean-field equations of the theory. 

The convexity of $e^{F}$ implies the existence of a unique local minimum 
of $F$. This connection is due to the fact that local minima of $F$ are 
inherited by $e^{F}$, so that the presence of {\em more than one} local 
minimum of $F$ would necessarily violate the convexity of $e^{F}$. The
exponential of the negative of the mean-field free energy given in Eq. 
(\ref{F}) can be expressed as:
\begin{equation}
\label{FGH}
  e^{F(\omega_{c},\chi_{c})} = \int  D\vec{x}(s) e^{-\frac{3}{2a_{p}^{2}}\int ds \dot{\vec{x}}^{2}(s)}
             {\cal G}(\omega_{c},\vec{x}(s)){\cal H}(\chi_{c},\vec{x}(s)) \, ,
\end{equation}
where
\begin{eqnarray}
\label{Gdef}
  {\cal G}(\omega_{c},\vec{x}(s))&\equiv&e^{\frac{\lambda}{2}\int \omega_{c}^{2}d\vec{r}
 -\lambda\int ds\omega_{c}(\vec{x}(s))} \, , \\
\label{Hdef}
   {\cal H}(\chi_{c},\vec{x}(s)) &\equiv& e^{-\frac{\beta\epsilon}{8\pi}\int \chi_{c}\Delta\chi_{c}d\vec{r}
 +c_{+}\int e^{e\beta\chi_{c}}d\vec{r} + c_{-}\int e^{-e\beta\chi_{c}}d\vec{r} -pe\beta\int ds\chi_{c}(\vec{x}(s))} \, .
\end{eqnarray}
(This can also be seen by inspection of Eqs. (1,2) with $\chi$, $\omega$ fixed 
at the saddle-point to ${-}i\chi_{c}$, ${-}i\omega_{c}$, respectively.)
Introducing the line distribution 
$j(\vec{r}){\equiv}{\int}ds\delta(\vec{r}{-}\vec{x}(s))$, 
Eqs. (\ref{Gdef},\ref{Hdef}) become
\begin{eqnarray}
\label{Gdef2}
  {\cal G}(\omega_{c},\vec{x}(s))&\equiv&e^{\frac{\lambda}{2}\int \omega_{c}^{2}d\vec{r}
 -\lambda\int \omega_{c}(\vec{r})j(\vec{r})d\vec{r}} \, , \\
\label{Hdef2}
   {\cal H}(\chi_{c},\vec{x}(s)) &\equiv& e^{-\frac{\beta\epsilon}{8\pi}\int \chi_{c}\Delta\chi_{c}d\vec{r}
 +c_{+}\int e^{e\beta\chi_{c}}d\vec{r} + c_{-}\int e^{-e\beta\chi_{c}}d\vec{r} -pe\beta\int \chi_{c}(\vec{r})j(\vec{r})d\vec{r}} \, .
\end{eqnarray}
  From Eq. (\ref{FGH}), it is apparent that the exponential of $F$ is a 
sum (over polymer configurations)  of positively weighted products of 
functionals of the fields $\omega_{c}$ and $\chi_{c}$ {\em separately}. 
Accordingly, the convexity of $e^{F}$ will be guaranteed once we can 
demonstrate the convexity of the functionals ${\cal G}$ (as a function of 
$\omega_{c}$) and $\cal{H}$ (as a function of $\chi_{c}$) for any arbitrary 
fixed polymer configuration $\vec{x}(s)$.

The second functional derivative of ${\cal G}$ is found after a short 
calculation to be
\begin{eqnarray}
 K_{G}(\vec{r},\vec{r}\,')&\equiv& \frac{\delta^{2}{\cal G}}{\delta\omega_{c}(\vec{r})\delta\omega_{c}(\vec{r}\,')} \nonumber \\
 &=& \left\{\lambda\delta(\vec{r}-\vec{r}\,')+\lambda^{2}\left[\omega_{c}(\vec{r})-j(\vec{r})\right]\left[\omega_{c}(\vec{r}\,')-j(\vec{r}\,')\right]\right\}{\cal G} \, .
	\label{Gpos}
\end{eqnarray}
A necessary and sufficient condition for a function $W(\vec{r},\vec{r}\,')$ to be positive semidefinite is:
\begin{equation}
\int d\vec{r}\int d\vec{r}\,'f(\vec{r})W(\vec{r},\vec{r}\,')f(\vec{r}\,')\,{\geq}\,0
	\label{eq.posdef}
\end{equation}
for any function $f(\vec{r})$.
The positivity of the kernel  $K_{G}$ in Eq. (\ref{Gpos}) is apparent  and 
can be demonstrated by verifying Eq. (\ref{eq.posdef}):
\begin{eqnarray}
&&\int d\vec{r}\int d\vec{r}\,'f(\vec{r})K_{G}(\vec{r},\vec{r}\,')f(\vec{r}\,')
\nonumber \\
&=&\lambda{\cal{G}}\int d\vec{r}f^{2}(\vec{r})+\lambda^{2}{\cal{G}}\left\{\int d\vec{r}f(\vec{r})\left[\omega_{c}(\vec{r})-j(\vec{r})\right]\right\}^{2}\,\geq\,0 \, .
	\label{posG}
\end{eqnarray}
The exponent 
of $\cal{H}$ can be written:
\begin{eqnarray}
&&-\frac{\beta\epsilon}{8\pi}\int \chi_{c}\Delta\chi_{c}d\vec{r}
 +c_{+}\int e^{e\beta\chi_{c}}d\vec{r} + c_{-}\int e^{-e\beta\chi_{c}}d\vec{r} -pe\beta\int \chi_{c}(\vec{r})j(\vec{r})d\vec{r} \nonumber \\
&=&-\frac{\beta\epsilon}{8\pi}\int \left(\chi_{c}\Delta\chi_{c}+\frac{8{\pi}pe}{\epsilon}\chi_{c}(\vec{r})j(\vec{r})\right)d\vec{r}+c_{+}\int e^{e\beta\chi_{c}}d\vec{r} + c_{-}\int e^{-e\beta\chi_{c}}d\vec{r} \nonumber \\
&=&-\frac{\beta\epsilon}{8\pi}\left[\int \left(\chi_{c}+\frac{1}{\Delta}\frac{4{\pi}pe}{\epsilon}j\right)\Delta\left(\chi_{c}+\frac{1}{\Delta}\frac{4{\pi}pe}{\epsilon}j\right)d\vec{r}-\left(\frac{4{\pi}pe}{\epsilon}\right)^{2}\int j\Delta^{-1}jd\vec{r}\right] \nonumber \\
&&+c_{+}\int e^{e\beta\chi_{c}}d\vec{r} + c_{-}\int e^{-e\beta\chi_{c}}d\vec{r} \nonumber \\
&=&-\frac{\beta\epsilon}{8\pi}\int \left(\chi_{c}+\hat{j}\right)\Delta\left(\chi_{c}+\hat{j}\right)d\vec{r}-\left(\frac{4{\pi}pe}{\epsilon}\right)^{2}\int j\Delta^{-1}jd\vec{r} \nonumber \\
&&+c_{+}\int e^{e\beta\chi_{c}}d\vec{r} + c_{-}\int e^{-e\beta\chi_{c}}d\vec{r} \, ,
\end{eqnarray}
where
\begin{equation}
\hat{j}\equiv\frac{1}{\Delta}\frac{4{\pi}pe}{\epsilon}j \, .
\end{equation}
Hence,
\begin{equation}
\frac{\delta\cal{H}}{\delta\chi_{c}(\vec{r})}=\left[-\frac{\beta\epsilon}{4\pi}\Delta\left(\chi_{c}+\hat{j}\right)(\vec{r})+c_{+}e{\beta}e^{e\beta\chi_{c}(\vec{r})}-c_{-}e{\beta}e^{-e\beta\chi_{c}(\vec{r})}\right]{\cal{H}} \, ,
\end{equation}
and therefore
\begin{eqnarray}
 K_{H}(\vec{r},\vec{r}\,')&\equiv&\frac{\delta^{2}{\cal H}}{\delta\chi_{c}(\vec{r})\delta\chi_{c}(\vec{r}\,')} \nonumber \\
&=&\left[-\frac{\beta\epsilon}{4\pi}\Delta\left(\chi_{c}+\hat{j}\right)(\vec{r})+c_{+}e{\beta}e^{e\beta\chi_{c}(\vec{r})}-c_{-}e{\beta}e^{-e\beta\chi_{c}(\vec{r})}\right] \nonumber \\ 
&&{\times}\left[-\frac{\beta\epsilon}{4\pi}\Delta\left(\chi_{c}+\hat{j}\right)(\vec{r}\,')+c_{+}e{\beta}e^{e\beta\chi_{c}(\vec{r}\,')}-c_{-}e{\beta}e^{-e\beta\chi_{c}(\vec{r}\,')}\right]\cal{H} \nonumber \\
&&-\frac{\beta\epsilon}{4\pi}\Delta_{\vec{r}}\delta(\vec{r}-\vec{r}\,')\cal{H} \nonumber \\
&&+e^{2}\beta^{2}\delta(\vec{r}-\vec{r}\,')\left(c_{+}e^{e\beta\chi_{c}(\vec{r})}+c_{-}e^{-e\beta\chi_{c}(\vec{r})}\right)\cal{H} \, .
	\label{Hpos}
\end{eqnarray}
Property (\ref{eq.posdef}) can be easily
verified for the first and third terms of Eq. (\ref{Hpos}).
For the second term one finds:
\begin{eqnarray}
{\int}d\vec{r}{\int}d\vec{r}\,'f(\vec{r})\left[-\Delta_{\vec{r}}\delta(\vec{r}-\vec{r}\,')\right]f(\vec{r}\,')\cal{H}&=&{\int}d\vec{r}f(\vec{r})\left[-{\Delta}f(\vec{r})\right]\cal{H} \nonumber \\
&=&{\int}d\vec{r}\left|\vec{\nabla}f\right|^{2}{\cal{H}}\, {\geq} \, 0 \, .
\end{eqnarray}
This establishes the positivity of the kernel $K_{H}$, which together with 
the positivity of $K_{G}$ proves the convexity of $e^{F}$, and consequently 
the existence of unique minimum of $F$, which itself guarantees uniqueness 
of the solution of the mean-field equations.

\newpage
\section{Three-dimensional lattice field theory results for a
charged polymer chain 
confined to two connected spheres of different radii}
We have used a recently developed LFT approach \cite{TCD} in order to 
obtain results of the solution of the mean-field
equations (\ref{eq:pb}) and (\ref{eq:nlse}), for 
polymer chains confined to 3D cavities. In this section we present results of 
the calculation of the equilibrium monomer distribution of a charged polymer 
chain constrained to move inside the volume of two connected spheres 
\cite{Park}.

After rescaling according to 
$f(\vec{r}){\rightarrow}{\beta}e\chi_{c}(\vec{r})$, 
$\Psi_{N}(\vec{r}){\rightarrow}a_{l}^{3/2}\Psi_{N}(\vec{r})$, and multiplying 
Eq. (\ref{eq:pb}) by $a_{l}^{3}$ ($a_l$ being the lattice spacing),
we solve the discretized version of equations 
(\ref{eq:pb}) and (\ref{eq:nlse}) on a 3D lattice: 
\begin{eqnarray}
\alpha\sum_{\vec{m}}\Delta_{\vec{n}\vec{m}}f_{\vec{m}}&=&\gamma_{+}e^{f_{\vec{n}}}-\gamma_{-}e^{-f_{\vec{n}}}-p\rho_{\vec{n}} \label{PBE} \\
\frac{a_{p}^{2}}{6a_{l}^{2}}\sum_{\vec{m}}\Delta_{\vec{n}\vec{m}}\Psi_{N,\vec{m}}&=&\frac{{\lambda}M}{a_{l}^{3}}\rho_{\vec{n}}\Psi_{N,\vec{n}}+pf_{\vec{n}}\Psi_{N,\vec{n}}-E_{N}\Psi_{N,\vec{n}} \label{SE} \, ,
\end{eqnarray}
where
\begin{eqnarray}
\alpha&=&\frac{{\varepsilon}a_{l}}{4\pi{\beta}e^{2}} \, , \\
\gamma_{\pm}&=&\frac{n_{\pm}}{\sum_{\vec{n}}e^{\pm f_{\vec{n}}}} \, ,
\end{eqnarray}
and the wavefunctions are dimensionless and normalized according to
\begin{equation}
\sum_{\vec{n}}\Psi_{N,\vec{n}}^{2}=1 \; ;  \; 
\end{equation}
thus, the density $\rho_{\vec{n}}$ sums to the total number of 
monomers, $M$.

Equations (\ref{PBE}) and (\ref{SE}) are solved simultaneously using the 
following relaxation procedure \cite{TCD}. First, the Schr\"odinger
Eq. (\ref{SE}) is solved 
for $f_{\vec{n}}{=}0$ and ignoring the
nonlinear (monomer repulsion) potential term. 
The resulting $\Psi_{N,\vec{n}}$'s and corresponding energy levels
$E_N$ (wavefunctions and energy eigenvalues of a particle confined
to a ``box" consisting of two fused spheres) are used 
to calculate $\rho_{\vec{n}}$, then the Poisson-Boltzmann
Eq. (\ref{PBE}) is solved at each lattice 
point using a simple line minimization procedure \cite{walsh}.
The process is repeated and the 
coefficients $\gamma_{\pm}$ are updated after 
a few iterations until a predetermined accuracy is achieved. Then 
the resulting $f_{\vec{n}}$ is used in Eq. (\ref{SE}), which is solved
using the Lanczos method \cite{lanc}
for a 
new set of $\Psi_{N,\vec{n}}$'s to be used in calculating
an updated version of the monomer density
$\rho_{\vec{n}}$. 
This density is then inputted into Eq. (\ref{PBE}) and a new
version of $f_{\vec n}$ is computed.
For numerical stability, the updated $f_{\vec n}$ inputted into
Eq. (\ref{SE}) is obtained 
by adding a small fraction of the new $f_{\vec{n}}$ (just obtained
from Eq. (\ref{PBE})) to the old one
(obtained 
from the previous iteration). The same ``slow charging" procedure is used for 
updating $\rho_{\vec{n}}$ in the nonlinear potential term of the Schr\"odinger 
equation (\ref{SE}).

The procedure is applied to the following two systems:

\hspace{3 mm}
(1) A polymer in a cavity consisting of two spheres of radii $R_{1}$ and 
$R_{2}$, carved out and sharing one common point on a lattice, the distance 
between the centers of the two spheres being $R_{1}{+}R_{2}$.

\hspace{3 mm}
(2) Same as (1), except that the spheres are now embedded in each other by 
one more lattice spacing, that is, the distance between their centers is now 
$R_{1}{+}R_{2}{-}a_{l}$.

We have used the following parameters in relative units: $R_{1}{=}1.0$, 
$R_{2}{=}0.8$, $a_{p}{=}0.2$, $\lambda{=}0.001$, and the two spheres are 
carved inside a cube of 44 lattice points on each side with $a_{l}{=}0.1$. 
In absolute units $a_{p}{=}5$\AA.

\begin{figure}[!]
\vspace{-0.2 cm}
\psfig{file=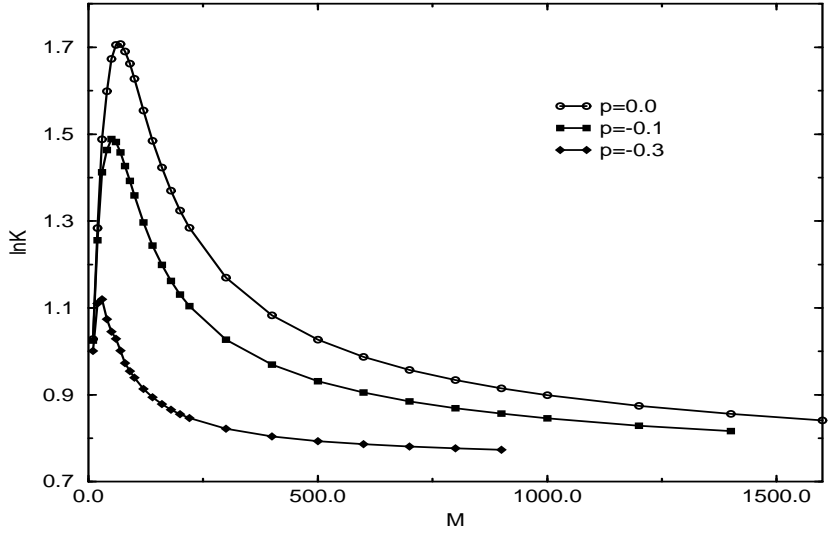,width=370pt,height=240pt,angle=270}
\caption{ The ratio ${\ln}K$ vs. $M$ for system (1) for varying monomer 
charge $p$ and fixed number of negative impurity ions $n{=}600$, which 
corresponds to a concentration $C{\approx}0.75M$.}
\end{figure}

\begin{figure}[!]
\psfig{file=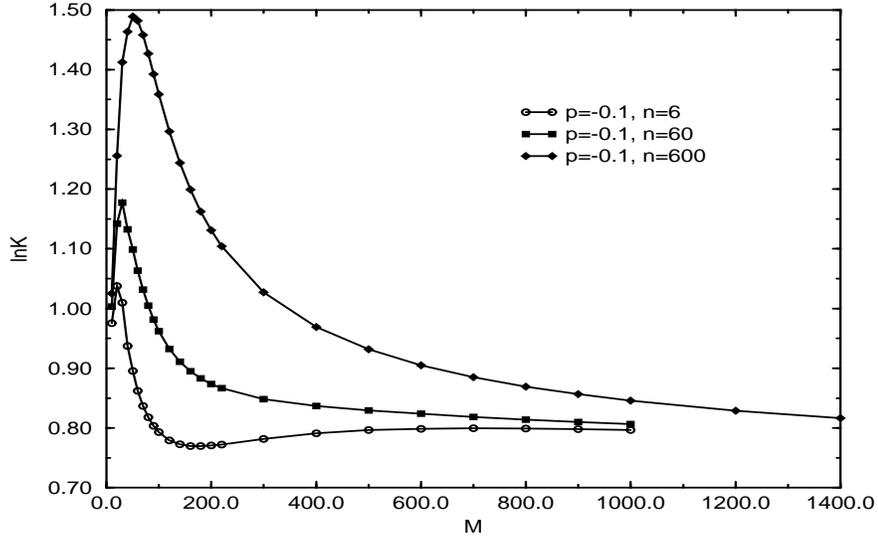,width=370pt,height=240pt,angle=270}
\caption{ The ratio ${\ln}K$ vs. $M$ for system (1) for varying number of 
negative impurity ions $n$ and fixed monomer charge $p$. The corresponding 
concentrations are $C(n{=}6){\approx}0.75{\times}10^{-2}M$, 
$C(n{=}60){\approx}0.75{\times}10^{-1}M$, and $C(n{=}600){\approx}0.75M$, 
respectively.}
\end{figure}

We have computed the partition coefficient 
$K{\equiv}\left<M_{1}\right>/\left<M_{2}\right>$ as a 
function of the total number of monomers in the system, 
$M{=}\left<M_{1}\right>{+}\left<M_{2}\right>$, 
for varying monomer charge $p$ and varying number of ions in the system. The 
results for system (1) are plotted in Figures 1 and 2, with $n$ being the 
number of negative impurity ions in the system, while the number of positive 
ions is adjusted so that electroneutrality is preserved. Similar to
what was found
in the case of neutral polymers \cite{Ts},
we see here that for small $M$, ${\ln}K$ increases sublinearly with 
$M$.  This is followed by a 
turnover region, after which ${\ln}K$ decreases with $M$,
and for very large $M$ 
approaches a limit bounded from
below by the log of the ratio of the volumes 
of the two spheres.

In Figure 1 we show results for varying monomer charge $p$, keeping the 
number of impurity ions fixed. It is apparent that smaller monomer charge 
favors larger ${\ln}K$, thus making polymer separation easier. In Figure 2 we 
vary the number of impurity ions, showing that a large number
of impurity ions leads to screening 
of the monomer
charges from each other and thus to a behavior resembling that of a 
neutral chain. We observe an interesting feature in the case of small number 
of ions: the ${\ln}K$ curve goes through a turnover, then through a minimum 
and a maximum, before approaching its asymptotic value at large $M$.

In Figures 3 and 4 we show the analogous results to those of Figures 1 and 
2, respectively, for system (2). The same basic behavior is observed,
with the partition coefficient being smaller for system (2) than for
system (1)
under identical conditions of polymer length, monomer charge and
impurity ion concentration.  This is because of the wider
conduit in system (2), which diminishes the isolation of one sphere
from the other, thus enabling the polymer to move more freely between
them.

\begin{figure}[h]
\psfig{file=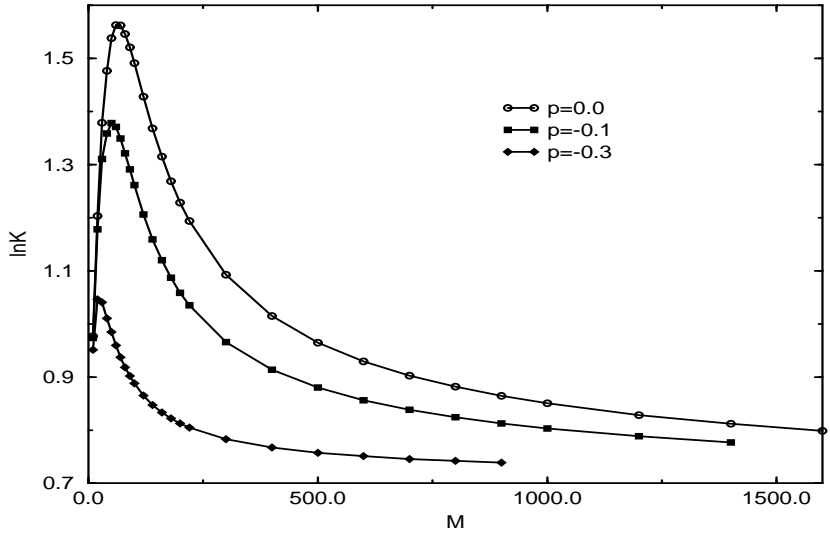,width=370pt,height=240pt,angle=270}
\caption{ The ratio ${\ln}K$ vs. $M$ for system (2) for varying monomer 
charge $p$ and fixed number of negative impurity ions $n{=}600$, which 
corresponds to a concentration $C{\approx}0.75M$.}
\end{figure}

\newpage
\begin{figure}[h]
\psfig{file=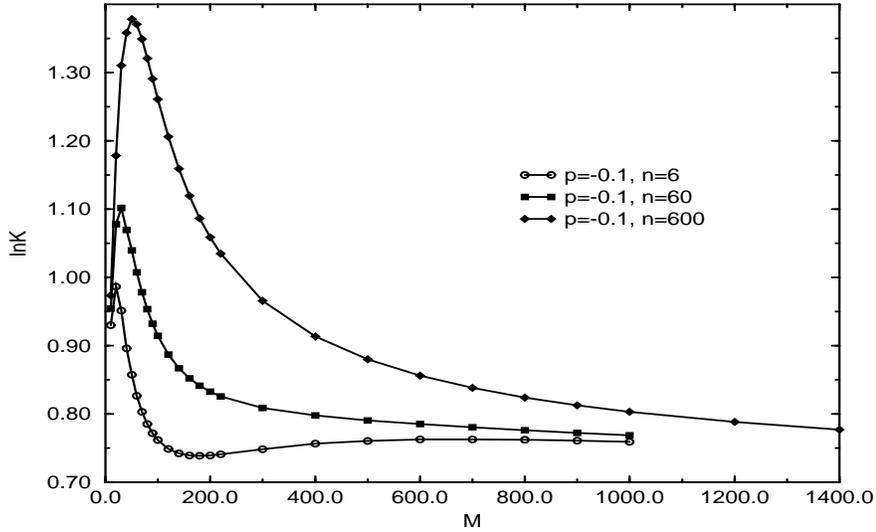,width=370pt,height=240pt,angle=270}
\caption{ The ratio ${\ln}K$ vs. $M$ for system (2) for varying number of 
negative impurity ions $n$ and fixed monomer charge $p$. The corresponding 
concentrations are $C(n{=}6){\approx}0.75{\times}10^{-2}M$, 
$C(n{=}60){\approx}0.75{\times}10^{-1}M$, and $C(n{=}600){\approx}0.75M$, 
respectively.}
\end{figure}

These results are consistent with the ideas presented in our 
previous work \cite{TCD}, namely, that higher impurity ion (electrolyte) 
concentrations
would favor better polymer separation between cavities of different size. 

\begin{figure}[!]
\psfig{file=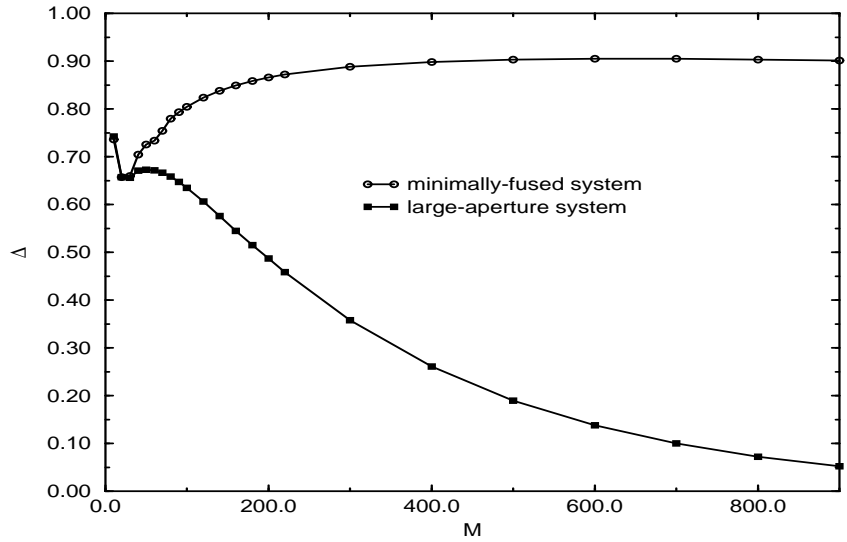,width=370pt,height=240pt,angle=270}
\caption{ The number 
$\Delta{\equiv}\exp\left[-M\left(E_{1}{-}E_{0}\right)\right]$ as a function 
of $M$ for system (1) (minimally-fused system) and system 
(2) (large-aperture system) for monomer charge $p{=}{-}0.3$ and number of 
negative impurity ions $n{=}600$.}
\end{figure}

Note that, as we found previously for an uncharged polymer 
chain \cite{Ts}, in the case of system (1) ground-state dominance 
of the polymer's Green's function does not set in
for any $M$. As we increase the total 
number of monomers, the first two energy levels come closer together and 
couple. Thus, even in the large $M$ limit we must retain both these states in 
order to calculate an accurate monomer density $\rho(\vec{r})$. In Figure 5 
we plot $\Delta{\equiv}\exp\left[-M\left(E_{1}{-}E_{0}\right)\right]$ vs. $M$ 
as an indicator of ground-state dominance ($\Delta{\rightarrow}0$ as 
$M{\rightarrow}\infty$) for $p{=}{-}0.3$ and $n{=}600$ for both systems (1) 
and (2). It is 
clear that in the case of system (1) ground-state dominance does not occur, 
while for system (2) it occurs, as would be expected for a polymer moving in 
a single cavity.

\newpage
\section{Conclusions}
We have extended the lattice field theory approach for the statistical 
mechanics of charged polymers in electrolyte solutions \cite{TCD} to the case 
where ground-state dominance fails in the polymer's Green's function. At the 
mean-field level all thermodynamic properties of the system are obtained 
from the solution of two coupled nonlinear equations. These equations involve 
the full monomer density,
which in general contains contributions from excited states in the
spectral decomposition of the polymer's Green's function.
We have also discussed 
the general shape of the negative of the system's total free energy 
functional, and have shown that it possesses a single minimum,
thus guaranteeing 
a unique solution of the mean-field equations.

We have used this approach to calculate the equilibrium partition coefficient 
$K$ of a Gaussian polymer chain in a system of two spheres connected by a 
narrow aperture, and have observed essentially
the same generic behavior seen in the case 
of excluded volume interactions only \cite{Ts}: the log of the partition 
coefficient increases sublinearly with the number of monomers in the chain, 
$M$, for small $M$, then it goes through a turnover region with a maximum, 
after which it decreases to an asymptotic value bounded from below by the log 
of the ratio of the volumes of the two spheres. Increasing the monomer 
charge makes this behavior less pronounced and reduces the
maximum in the ${\ln}K$ vs. $M$ 
curve, while increasing the impurity ion concentration leads to screening of 
the monomer charges, and makes the ${\ln}K$ vs. $M$ curve similar to the one 
in the case of a neutral polymer with excluded volume interactions. This 
supports our previous contention \cite{TCD} that higher impurity 
concentrations would lead to better polymer separation between cavities of 
different size, such as the cavities in a gel used to observe the ``entropic 
trapping'' phenomenon \cite{Asher}. 

The present
work demonstrates the failure of the notion of ground-state dominance 
in the case of a very narrow conduit between the two spheres.  For such
a system,
even in the large $M$ limit the first excited state of the polymer's Green's 
function must be retained in order to obtain an accurate mean-field
solution to the 
problem.

Of course, the accuracy of the venerable mean-field approximation 
\cite{Edwards,deG} for the systems considered here remains an outstanding 
issue. In the case of electrically neutral polymers, Monte-Carlo simulations 
may provide valuable benchmarks against which mean-field predictions can be 
compared. Such computations are currently under way.

\end{document}